\documentclass[preprint,12pt]{elsarticle}

\usepackage{amssymb}
\usepackage{graphicx}

\journal{Journal of Alloys and Compounds}

\begin{document}

\author[INT]{M. J. Winiarski}
\author[PWR]{P. Scharoch}
\author[PWR]{M. P. Polak}
\address[INT]{Institute of Low Temperature and Structure Research, Polish Academy of Sciences, Ok\'olna 2, 50-422 Wroc\l aw, Poland, EU}
\address[PWR]{Institute of Physics, Wroclaw University of Technology, Wyb. Wyspianskiego 27, 50-370 Wroclaw, Poland, EU}

\title{First principles prediction of structural and electronic properties of Tl$_x$In$_{1-x}$N alloy}

\begin{abstract}
Structural and electronic properties of zinc blende Tl$_x$In$_{1-x}$N alloy have been evaluated from first principles. The band structures have been obtained within the density functional theory (DFT), the modified Becke-Johnson (MBJLDA) approach for the exchange-correlation potential, and fully relativistic pseudopotentials. The calculated band-gap dependence on Tl content in this hypothetical alloy exhibits a linear behaviour up to the 25 \% of thalium content where its values become close to zero. In turn, the split-off energy at the $\Gamma$ point of the Brillouin zone, related to the spin-orbit coupling, is predicted to be comparable in value with the band-gap for relatively low thalium contents of about 5 \%. These findings suggest Tl$_x$In$_{1-x}$N alloy as a promising material for optoelectronic applications. Furthermore, the band structure of TlN reveals some specific properties exhibited by topological insulators.
\end{abstract}

\maketitle

\section{Introduction}

Group III metal nitrides have been widely explored materials for optoelectronic applications. Forming of ternary and quaternary alloys in the wurtzite phase offers a promising perspective of tuning their direct band-gaps starting from 0.65-0.69 eV \cite{PhysStatSolB_229_R1,APL_83_4963,JApplPhys_94_4457} for InN, through 3.50-3.51 eV \cite{JApplPhys_83_1429, JApplPhys_94_3675} for GaN, up to 6.1 eV for AlN \cite{JApplPhys_94_3675}, which is crucial for optoelectronics. Especially the In-based systems have been extensively investigated due to their small band-gaps \cite{PRB_80_075202,PhysStatSolB_249_485,ApplPhysLett_96_021208,PhysLettA_374_4767}.

Optical properties of some thalium III-V alloys have already been reported. First, the theoretical investigations of Tl$_x$In$_{1-x}$Sb \cite{ApplPhysLett_62_1857} and Tl$_{1-x}$In$_x$P \cite{ApplPhysLett_65_2714} suggested these materials as potential candidates for infrared detectors. Next, experimental studies extended this idea on ternary Tl$_{1-x}$In$_x$P \cite{JApplPhys_81_1704}, Tl$_{1-x}$Ga$_x$As, and Tl$_{1-x}$In$_x$As alloys \cite{PhysRevB_72_125209, JCrystGrow_237_1495}.

Although the synthesis of thalium nitride has not been reported yet, this compound has been studied theoretically \cite{MatSciEngB_103_258,JCrystGrow_281_151,CompMatSci_50_203} because of the predicted existence of semimetallic phase, i.e. the situation where creating an alloy by adding a wide band-gap component and controlling its content may cover a very broad range of band-gaps. The structural and elastic properties of TlN, as technologically important, have been investigated in equilibrium conditions and under hydrostatic pressure \cite{ChinPhysLett_27_080505, ChinPhysLett_28_100503}. The electronic and structural properties of Tl$_x$Al$_{1-x}$N \cite{ApplPhysLett_92_121914, MatSciSemProc_15_499} and Tl$_x$Ga$_{1-x}$N \cite{MatSciEngB_162_26} have been also studied \emph{ab initio}, however, these investigations were performed with the standard DFT exchange-correlation functionals.

The standard (LDA/GGA) DFT calculations lead to unrealistic band structures for narrow band-gap semiconductors, and as a result, to false conclusions about the real value of the band-gap as well as the spin-orbit coupling (SOC) driven separation between the heavy hole and the split-off band at the $\Gamma$ point of the Brillouin zone. Thus, such technologically important properties of semiconducting alloys as band-gap bowing and the split-off energy ($\Delta_{SO}$) cannot be studied within LDA/GGA approach.

In this work, the modified Becke-Johnson potential (MBJLDA) proposed by Tran and Blaha \cite{PRL_102_226401} was used for the band structure calculations. The performance of this method has been already tested for small band-gap systems, eg. InAs and InSb \cite{PRB_82_205212}, thus the results for zinc blende Tl$_x$In$_{1-x}$N alloy reported here may be considered as a reliable prediction of the band structure properties of this hypothetical novel alloy.

\section{Computational details}

The electronic structure calculations for zinc blende (ZB) Tl$_x$In$_{1-x}$N alloy have been performed with the use of the ABINIT code \cite{Abinit1, Abinit2}. First, the equilibrium geometries were found via stresses/forces relaxation for PAW atomic datasets generated with the Atompaw package \cite{Atompaw} with the Perdew-Wang \cite{LDA} parametrization for the exchange-correlation energy. This was followed by the calculations of the band structures with the use of the norm-conserving fully relativistic pseudopotentials generated using the Atomic Pseudopotential Engine \cite{APE}. The MBJLDA (TB09) \cite{PRL_102_226401} functional, which consists of the modified Becke-Johnson exchange potential combined with an LDA correlation, was used. This complex calculation scheme was necessary to obtain reasonable results of LDA predicted structural parameters and MBJLDA band-gaps with spin-orbit coupling effects included.
A supercell consisting of 16-atoms ($2\times2\times2$ multiplicity of primitive FCC cell) was used to simulate the alloy. The most possibly uniform arrangements of atoms were chosen. The total energy convergence in the plane wave basis was found to be sufficient for 30 Ha energy cutoff and the $4\times4\times4$  {\bf k}-point mesh with standard shifts for FCC lattice.

It should be noted that calculations in the pseudopotential approach lead to an underestimation of the $c$ parameter in the MBJLDA exchange-correlation functional \cite{PRL_102_226401}, and as a result, to a relatively low value of InN band-gap. In this work, $c$ = 1.335 was chosen to obtain the band-gap value of 0.69 eV, the same as computed with the full-potential MBJLDA calculations with the Wien2k code \cite{Wien2k}. An analogous discussion has been also contained in ref. \cite{InGaN_alchemia}.

\section{Results and discussion}

Equilibrium lattice parameters, split-off energies and band-gaps calculated for InN and TlN are gathered in Tab.\ref{table1}. Although the (LDA) lattice parameter obtained here for TlN is in a good agreement with the literature data, the $a$ = 4.948 \AA \ for InN is significantly higher than the value of 4.800 \AA  \ \cite{JPhysCondMat_14_9579} reported earlier, and slightly different than our former result \cite{InGaN_alchemia} from pseudopotential calculations. It is worth to note that the PAW atomic data used here contained the full set of  valence states, including the semicore In 4s and 4p states, thus the presented here structural parameters seem to be the most adequate DFT-derived values.

The MBJLDA band-gap value for ZB InN, E$_g$ = 0.69 eV, computed with the full-potential method (Wien2k), being reasonable compared to literature, was used as a reference value in fully relativistic pseudopotential calculations performed in this work. Interestingly, both the fully relativistic LDA and MBJLDA approaches gave a nonzero direct band-gap for TlN in the $\Gamma$ point, in contrast to former LDA/GGA investigations without SOC \cite{MatSciEngB_162_26}, however significantly smaller than 0.050 eV reported for fully relativistic GGA calculations \cite{ApplPhysLett_92_121914}.

\begin{figure}
\includegraphics[width=8cm]{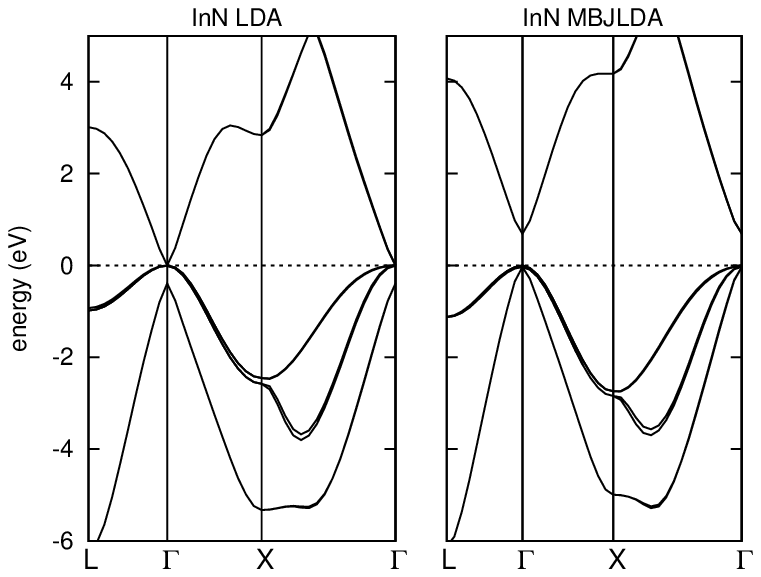}
\caption{Band structure of zinc blende InN, calculated within LDA and MBJLDA.}
\label{Fig1}
\end{figure}

The $\Delta_{SO}$ obtained with the LDA approach for InN is strongly overestimated. Since LDA calculations lead to artificially negative band-gap of InN, the position of the split-off band is not related to the SOC effects. The MBJLDA calculations provide  more proper description of InN band structure, presented in Fig. \ref{Fig1}. In this case, the real SOC-driven $\Delta_{SO}$ = 0.033 eV is an order of magnitude lower than 0.393 eV obtained with LDA, thus hardly visible in the scale of Fig. \ref{Fig1}, whereas some distinct splittings of heavy and light-hole bands can be seen in the $X$ point as well as along $X$-$\Gamma$ direction in the BZ of InN.

\begin{figure}
\includegraphics[width=8cm]{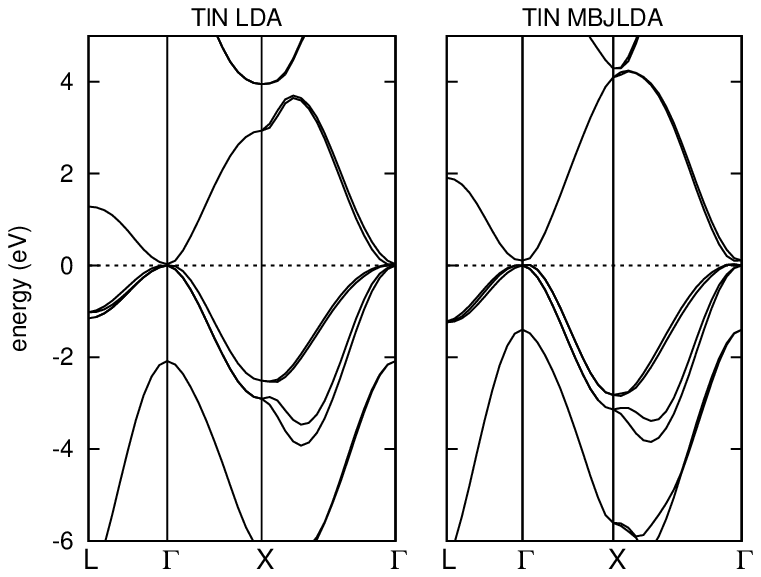}
\caption{Band structure of zinc blende TlN, calculated within LDA and MBJLDA.}
\label{Fig2}
\end{figure}

\begin{figure}
\includegraphics[width=8cm]{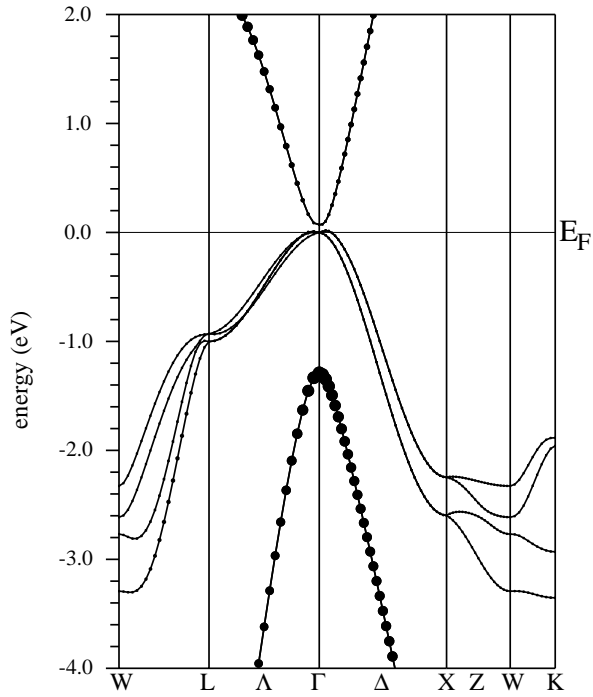}
\caption{Orbital band characters (fat bands) for s-electrons of Tl atom in zinc blende TlN indicating the s-type character of split-off band
(the minimum of the conduction band is dominated by p-electrons of N atom).}
\label{Fig3}
\end{figure}

One can consider an analogous issue for TlN, despite that in this compound the SOC effect could have been expected to be very strong. In this case the MBJLDA calculations also lead to a significant reduction of $\Delta_{SO}$ from 2.083 eV obtained with LDA to a more reasonable value of 1.397 eV. The band structure of TlN is depicted in Fig. \ref{Fig2}. The strong SOC effects are clearly visible in all directions in TlN band structure. A former study reported a band-gap inversion in this system \cite{MatSciEngB_162_26}, which is also present in the fully relativistic MBJLDA results, as can be seen in band character plot presented in Fig. \ref{Fig3}.

Interestingly, both LDA and MBJLDA approaches suggest a narrow and somewhat indirect band-gap in the vicinty of the $\Gamma$ point in BZ of TlN. This effect is related to the anomalous order of N 2p bands, since the quartet N 2p$_{3/2}$ is below the doublet N 2p$_{1/2}$. Analogous properties of band structure have been revealed for $\beta$-HgS which is suggested to be a strong 3D topological insulator \cite{PRL_106_236806}, thus TlN may also be considered as a candidate compound for a topological insulator.
Despite that the parity criteria \cite{topological} do not apply to the binary ZB-type structure due to the lack of inversion symmetry, the simple predictions based only on a topological band-order are generaly accepted in literature, e.g. for InSb \cite{PRB_85_195114}. Therefore, in this work TlN is expected to exhibit a similar surface state behaviour to that of $\beta$-HgS, although the surface calculations for TlN have not been performed.

\begin{figure}
\includegraphics[width=8cm]{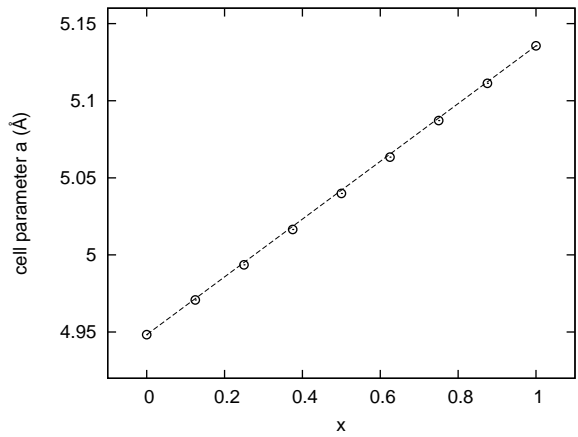}
\caption{LDA calculated lattice parameters of zinc blende Tl$_x$In$_{1-x}$N alloy as a function of thalium content.}
\label{Fig4}
\end{figure}

The lattice parameter of ZB Tl$_x$In$_{1-x}$N alloy reveals Vegard's law behaviour, as depicted in Fig.\ref{Fig4}. A similar linear dependence of lattice parameters of an alloy was reported for Tl$_x$Al$_{1-x}$N \cite{ApplPhysLett_92_121914}, which may suggest that the deviation from Vegard's law for Tl$_x$Ga$_{1-x}$N in ref. \cite{MatSciEngB_162_26} needs a careful verification.

\begin{figure}
\includegraphics[width=8cm]{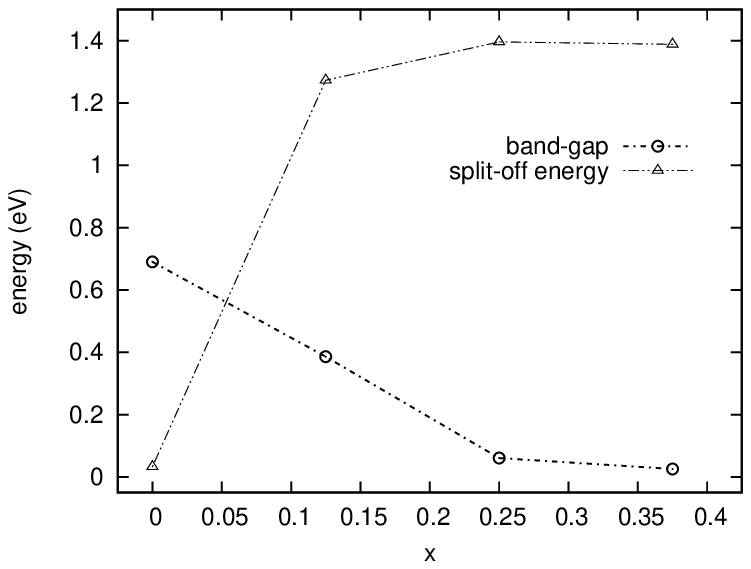}
\caption{MBJLDA calculated band-gap and split-off energy for Tl$_x$In$_{1-x}$N alloy.}
\label{Fig5}
\end{figure}

The band-gaps (E$_g$), and the split-off energies ($\Delta_{SO}$) of Tl$_x$In$_{1-x}$N alloy are depicted in Fig.\ref{Fig5}. The dependence of the alloy band-gap as a function of Tl content predicted here is somewhat unusual compared to those of other group III-V nitride alloys. However, a similar behaviour has also been reported for Tl$_x$Al$_{1-x}$N for a relatively high thalium contents \cite{MatSciSemProc_15_499}. This effect is related to the semimetallic character of TlN and its strongly ionic bonding \cite{MatSciEngB_103_258}, on the contrary to semiconducting systems. Thus, the introduction of Tl atoms into InN leads to a strong modification of electronic structure.

The calculated $\Delta_{SO}$ for Tl$_x$In$_{1-x}$N alloy, presented in Fig.\ref{Fig5}, indicates a very strong enhancement of
SOC introduced by Tl atoms. Since InN is a small band-gap system, the predicted value of $\Delta_{SO}$ in Tl$_x$In$_{1-x}$N alloy
is comparable to the band-gap for a relatively low thalium contents, about 5\%. The $\Delta_{SO}$ is an important parameter for
optoelectronic devices since it is connected with the unwanted non-radiative  Auger recombination \cite{PRL_110_177406} (CHSH transition), which is suppressed when $\Delta_{SO}$
is greater in value than E$_g$. E.g. in the Bi-dopped GaAs this situation takes place for Bi content greater than 
 9-10 \%  \cite{PhysRevB_87_115104}.
Therefore, the presented here results for Tl$_x$In$_{1-x}$N alloy indicate that this system is a promising material for applications in
optoelectronics, even for a relatively low thalium contents. However, despite the successful synthesis of
Tl$_{1-x}$In$_x$P \cite{JApplPhys_81_1704}, Tl$_{1-x}$Ga$_x$As, and Tl$_{1-x}$In$_x$As alloys \cite{PhysRevB_72_125209, JCrystGrow_237_1495}
there are still no reports on experimental data for any Tl-dopped group III nitrides.

In turn, thalium-dopped InN systems may be considered as candidate compounds for topological insulators with relatively wide band-gaps, thus the topological band-order in Tl$_x$In$_{1-x}$N alloy seems to be an interesting topic for further investigations.

\section{Conclusions}

The fully relativistic MBJLDA investigations of electronic structure of Tl$_x$In$_{1-x}$N alloy predict a linear decrease in alloy band-gap with increasing thalium content, as well as the semimetallic character above $x \approx 0.25$. The high split-off energy in the $\Gamma$ point of the BZ, introduced by spin-orbit coupling, indicates that this system is a promising material for optoelectronic devices. Futhermore, the Tl$_x$In$_{1-x}$N alloy may also exhibit a topological insulator behaviour. The results presented in this work should encourage further theoretical and experimental studies of Tl-dopped group III nitride alloys.

\section*{Acknowledgments}
The calculations were performed in Wroc\l aw Centre for Networking and Supercomputing.

\begin{table}
\caption{Lattice parameter {\it a}, band-gap E$_g$, split-off energy ($\Delta_{SO}$) of zinc blende InN and TlN}
\label{table1}
\begin{tabular}{llll}
 &  {\it a} (\AA) &  E$_g$ (eV) & $\Delta_{SO}$ (eV) \\ \hline
InN: & & &  \\
LDA & 4.948 & 0.00 &  0.393 \\
MBJLDA & - & 0.69 &  0.033 \\
ref \cite{InGaN_alchemia} & 4.970 & 0.64 & - \\
ref \cite{JPhysCondMat_14_9579} & 4.801 & 0.75 & -\\
ref \cite{SSC_116_421} & - & - & 0.013 \\
TlN: & & &  \\
LDA & 5.136 & 0.03 & 2.083 \\
MBJLDA & - & 0.11 & 1.397 \\
ref\cite{JCrystGrow_281_151} & 5.139 & - & - \\
ref\cite{MatSciEngB_162_26} & 5.133 & - & 2.000 \\
\end{tabular}
\end{table}

\begin{thebibliography}{50}

\bibitem{PhysStatSolB_229_R1} V. Y. Davydov, Phys. Stat. Sol. B 229 (2002) R1.
\bibitem{APL_83_4963} S. X. Li, J. Wu, E. E. Haller, W. Walukiewicz, W. Shan, H. Lu, W. J. Schaff, Appl. Phys. Lett. 83 (2003) 4963.
\bibitem{JApplPhys_94_4457} J. Wu, W. Walukiewicz, W. Shan, K. M. Yu, J. W. Ager, S. X. Li, E. E. Haller, H. Lu, W. J. Schaff, J. Appl. Phys. 94 (2003) 4457.
\bibitem{JApplPhys_83_1429} Y. C. Yeo, T. C. Chong, M. F. Li, J. Appl. Phys. 83 (1998) 1429.
\bibitem{JApplPhys_94_3675} I. I. Vurgaftman, J. R. Meyer, J. Appl. Phys. 94 (2003) 3675.

\bibitem{PRB_80_075202} I. Gorczyca, S. P. Lepkowski, T. Suski, N. E. Christensen, A. Svane, Phys. Rev. B 80 (2009) 075202.
\bibitem{PhysStatSolB_249_485} E. Sakalauskas, O. Tuna, A. Kraus, H. Bremers, U. Rossow, C. Giesen, M. Heuken, A. Hangleiter, G. Gobsch, R Goldhahn, Phys. Stat. Sol. B 249 (2012) 485.
\bibitem{ApplPhysLett_96_021208} P. G. Moses, C. G. Van de Walle, Appl. Phys. Lett. 96 (2010) 021208.
\bibitem{PhysLettA_374_4767} S. Zhang, J. Shi, S. Zhu, F. Wang, M. Yang, Z. Bao, Phys. Lett. A 374 (2010) 4767.

\bibitem{ApplPhysLett_62_1857} M. Van Schilfgaarde, A. Sher, A. B. Chen, Appl. Phys. Lett. 62 (1993) 1857.
\bibitem{ApplPhysLett_65_2714} M. Van Schilfgaarde, A. B. Chen, S. Krishnamurthy, A. Sher, Appl. Phys. Lett. 65 (1994) 2714.
\bibitem{JApplPhys_81_1704} K. Yamamoto, H. Asahi, M. Fushida, K. Iwata, S. Gonda, J. Appl.Phys. 81 (1997) 1704.
\bibitem{PhysRevB_72_125209} R. Beneyton, G. Grenet, Ph. Regreny, M. Gendry, G. Hollinger, B. Canut, C. Priester, Phys. Rev. B 72 (2005) 125209.
\bibitem{JCrystGrow_237_1495} Y. Kajikawa, H. Kubota, S. Asahina, N. Kanayama, J. Cryst. Growth 237 (2002) 1495.

\bibitem{MatSciEngB_103_258} A. Zaoui, Mat. Sci. Eng. B 103 (2003) 258.
\bibitem{JCrystGrow_281_151} A. Ferreira da Silva, A. N. Souza Dantas, J. S. de Almeida, R. Ahuja, C. Perssone, J. Cryst. Grow. 281 (2005) 151.
\bibitem{CompMatSci_50_203} L. Shi, Y. Duan, L. Qin, Comput. Mater. Sci. 50 (2010) 203
\bibitem{ChinPhysLett_27_080505} L.-W. Shi, Y.-F. Duan, L.-X. Qin, Chin.Phys.Lett. 27 (2010) 0880505.
\bibitem{ChinPhysLett_28_100503} L.-W. Shi, Y.-F. Duan. Y.-F. Yang, L.-X. Tang, Chin. Phys. Lett. 28 (2011) 100503.
\bibitem{ApplPhysLett_92_121914} N. Souza Dantas, J. S. de Almeida, R. Ahuja, C. Persson, A. Ferreira da Silva, Appl. Phys. Lett 92 (2008) 121914.
\bibitem{MatSciSemProc_15_499} L. Shi, Y. Duan, X. Yang, G. Tang, L. Qin, L. Qiu, Matter. Sci. Semicond. Process. 15 (2012) 499.
\bibitem{MatSciEngB_162_26} N. Saidi-Houat, A. Zaoui, A. Belabbes, M. Ferhat, Mat. Sci. Eng. B 162 (2009) 26.

\bibitem{PRL_102_226401} F. Tran, P. Blaha, Phys. Rev. Lett 102 (2009) 226401.
\bibitem{PRB_82_205212} Y.-S. Kim, M. Marsman, G. Kresse. F. Tran, P. Blaha, Phys. Rev. B 82 (2010) 205212.
\bibitem{Abinit1} X. Gonze et al., Comput. Mater. Sci. 25 (2002) 478.
\bibitem{Abinit2} X. Gonze et al., Comput. Phys. Commun. 180 (2009) 2582.
\bibitem{Atompaw} A. R. Tackett, N. A. W. Holzwarth, G. E. Matthews, Comput. Phys. Commun. 135 (2001) 329.
\bibitem{LDA} J. P. Perdew, Y. Wang, Phys. Rev. B 45 (1992) 13244.
\bibitem{APE} M. Oliveira, F. Nogueira, Comput. Phys. Comm. 178 (2008) 524.
\bibitem{Wien2k}  P. Blaha, K. Schwarz, P. Sorantin, S. B. Trickey, Comp. Phys. Commun. 59 (1990) 399.

\bibitem{InGaN_alchemia} P. Scharoch, M. J. Winiarski, M. P. Polak, Comput. Mater. Sci. 81 (2014) 358.
\bibitem{JPhysCondMat_14_9579} S. Q. Wang, H. Q. Ye, J. Phys.: Condens. Matter 14 (2002) 9579.
\bibitem{SSC_116_421} M. Cardona, N.E. Christensen, Solid State Commun. 116 (2000) 421.

\bibitem{PRL_106_236806} F. Virot, R. Hayn, M. Richter, J. van den Brink, Phys. Rev. Lett. 106 (2011) 236806.
\bibitem{topological} L. Fu, C. L. Kane, Physical Review B 76 (2007) 045302.
\bibitem{PRB_85_195114} W. Feng, W. Zhu, H. H. Weitering, G. M. Stocks, Y. Yao, D. Xiao, Phys. Rev. B 85 (2012) 195114.

\bibitem{PRL_110_177406} J. Iveland, L, Martinelli, J. Peretti, J.S. Speck, and C. Weisbuch, Phys. Rev. Lett. 110 (2013) 177406.

\bibitem{JApplPhys_111_113108} Z. Batool, K. Hild, T. J. C. Hosea, X. Lu, T. Tiedje, S. J. Sweeney, J. Appl. Phys. 111 (2012) 113108.
\bibitem{PhysRevB_87_115104} M. Usman, C. A. Broderick, Z. Batool, K. Hild, T. J. C. Hosea, S. J. Sweeney, E. P. O'Reilly, Phys. Rev. B 87 (2013) 115104.

\end{thebibliography}
\end{document}